# Robust R-Peak Detection in Low-Quality Holter ECGs using 1D Convolutional Neural Network

Muhammad Uzair Zahid, Serkan Kiranyaz, Turker Ince, Ozer Can Devecioglu, Muhammad E. H. Chowdhury, Amith Khandakar, Anas Tahir and Moncef Gabbouj

*Abstract— Objective:* Noise and low quality of ECG signals acquired from Holter or wearable devices deteriorate the accuracy and robustness of R-peak detection algorithms. This paper presents a generic and robust system for R-peak detection in Holter ECG signals. While many proposed algorithms have successfully addressed the problem of ECG R-peak detection, there is still a notable gap in the performance of these detectors on such low-quality ECG records. *Methods:* In this study, a novel implementation of the 1D Convolutional Neural Network (CNN) is used integrated with a verification model to reduce the number of false alarms. This CNN architecture consists of an encoder block and a corresponding decoder block followed by a sample-wise classification layer to construct the 1D segmentation map of R-peaks from the input ECG signal. Once the proposed model has been trained, it can solely be used to detect R-peaks possibly in a single channel ECG data stream quickly and accurately, or alternatively, such a solution can be conveniently employed for real-time monitoring on a lightweight portable device. *Results:* The model is tested on two open-access ECG databases: The China Physiological Signal Challenge (2020) database (CPSC-DB) with more than one million beats, and the commonly used MIT-BIH Arrhythmia Database (MIT-DB). Experimental results demonstrate that the proposed systematic approach achieves 99.30% F1-score, 99.69% recall, and 98.91% precision in CPSC-DB, which is the best R-peak detection performance ever achieved. Results also demonstrate similar or better performance than most competing algorithms on MIT-DB with 99.83% F1-score, 99.85% recall, and 99.82% precision. *Significance:* Compared to all competing methods, the proposed approach can reduce the false-positives and false-negatives in Holter ECG signals by more than 54% and 82%, respectively. *Conclusion:* Finally, the simple and invariant nature of the parameters leads to a highly generic system and therefore applicable to any ECG dataset.

M. U. Zahid, S. Kiranyaz, M. Chowdhury, A. Khandakar, A. Tahir are with Electrical Engineering, College of Engineering, Qatar University, Qatar; e-mail: m.uzair@qu.edu.qa, mkiranyaz@qu.edu.qa.

T. Ince is with the Electrical & Electronics Engineering Department, Izmir University of Economics, Turkey; e-mail: turker.ince@izmirekonomi.edu.tr.

M. Gabbouj and O. C. Devecioglu are with the Department of Computing Sciences, Tampere University, Finland; e-mail: Moncef.gabbouj@tuni.fi.

## I. INTRODUCTION

ACCURATE detection of R-peaks is essential for the diagnosis of cardiovascular diseases (CVD) in electrocardiogram (ECG) ) signals. The QRS complex, which is dependent on the accurate detection of R-peak, is the most important feature in the diagnosis of several cardiac pathologies. Although a wide variety of modalities such as blood tests, stress tests, echocardiograms, and chest X-rays have been used for CVD diagnosis; Electrocardiography is perhaps the most significant process for non-invasively monitoring and clinical diagnosis. ECG records the time evolution of the heart's electrical activity and helps in the diagnosis of numerous cardiovascular abnormalities such as premature ventricular contraction (PVC or V beats) and supraventricular premature beats (SPB or S beats). The introduction of low-cost wearable ECG monitors gives us a significant motive to investigate highly accurate and robust automated detection of R-peaks in single-lead ECG signals.

Although many R-peak detection methods have been proposed throughout the last several decades, robust and accurate peak detection is still a challenging problem, especially in noisy, degraded, and dynamically varying rhythms, particularly common in Holter registers. Holter monitor is an ambulatory ECG device, for portable cardiac monitoring and frequently, it is corrupted by a substantial proportion of motion artifacts [1]. Thus, R-peak detection is severely affected by the ECG signals with such poor signal quality and high noise levels [2].

R-peak detection (segmentation) is the base of arrhythmia detection and classification. ECG-based applications are generally divided into four phases: preprocessing (filtering), ECG signal segmentation (QRS complex detection), feature extraction, and classification algorithms. Poor segmentation performance propagates the error to subsequent steps and directly reduces classification efficiency. Much of the work in the literature focuses on minimizing the number of false positives during the classification step, ignoring the fact that the error started to spread during the segmentation step [3], [4].

One of the most widely used R-peak detection algorithms was developed by Pan and Tompkins (P&T)[5], which served as the benchmark for more than three decades. Hamilton algorithm emerged as a modification to the classical P&T method based on an optimized decision rule process [6]. There are several other popular methods for R-peak detection based on various signal processing techniques such as Wavelet transform [7], Hilbert transform [8] as well as their modifications with improved detection thresholds [9], Phasor transform [10], and ensemble empirical mode decomposition [11], [12]. Discrete wavelet transform (DWT) decomposes a signal into different frequency components, each with distinct coefficients, which contain sufficient information of the original signal [13]. Sahoo et al. have reported 99.87 % sensitivity in QRS complex detection using DWT [14]. Similarly, other techniques such as empirical mode decomposition (EMD) have helped in removing baseline wandering in ECG signals [15], Kabir and Shahnaz have provided better time resolution in removing noise from ECG using the combination of EMD and DWT [16]. Digital filters along with EMD have also been used to improve R-peak detection [17]. Slimange and Ali in [12] have used a nonlinear transformation technique based on EMD to achieve 99% sensitivity and specificity. Hilbert and wavelet transform with varying thresholds have also been used for R-peak detection [18].

Many of the aforementioned algorithms depend on two main stages: R-peak enhancement and detection. The first stage involves preprocessing techniques such as filter banks and spectral analysis, which are sometimes referred to as feature extraction to enhance the R-peak as compared to other ECG waves (P and T waves). The detection stage involves decision making based on a threshold to define the onset and offset of the R-wave. Such algorithms are lightweight and can be used easily with wearable or embedded devices. However, they usually perform well only for high quality and clean ECG signals and are not robust to noise [19], [20]. Thus, many proposed algorithms evaluate their detectors on MIT-BIH arrhythmia or similar datasets that contain high-quality ECG signals in standard clinical settings. Most of the QRS detection algorithms had high detection sensitivity and positive predictivity in the MIT-BIH arrhythmia data set. (>99%) [21], [22]. The performance of such algorithms significantly deteriorates when tested in a highly dynamic and noisy ECG dataset with severe artifacts. Even the basic QRS detection can be invalid in the low signal quality ECG analysis. There are some public datasets [23], [24] that contain noisy ECG signals with R-peak annotations, but they contain a limited number of ECG beats, and hence they are not suitable for a proper performance evaluation in general.

The applications of machine learning, e.g., sigmoid radial basis functions [25], Hidden Markov model (HMM) [26], and artificial neural network-based QRS analysis [27] have also been investigated for peak detection and classification of ECG signals. Rodriguez et al. have proposed a novel QRS complex detection method [28] using adaptive threshold with Hilbert transform and Principal component analysis (PCA). Deep learning has been very successful in speech recognition, natural language processing, and computer vision. In recent years, one-dimensional convolutional neural network (1D-CNN) has also been intensively studied for its speed and efficiency in managing complex tasks, that have been demonstrated in several signal processing applications [29], [30] motor fault detection [31] and classification of electrocardiogram signals [32] and advance warning system for cardiac arrhythmias [33]. Two parallel 1D residual neural networks were proposed by Wang *et al.*, which can obtain the time domain characteristics of QRS waveform and attained 99.98% positive predictive value and 99.92% sensitivity on MIT-BIH Arrhythmia dataset (MIT-DB)[34]. Laitala et al. have proposed R-peak detection using Long Short-Term Memory (LSTM) network which excels at temporal modeling tasks that include long-term dependencies, making it suitable for ECG analysis [35]. Similarly, Vijayrangan et al. have proposed a deep learning-based method using U-Net combined with Inception and Residual blocks on a combination of datasets and have achieved around 98.37 % accuracy [36]. A two-level CNN was proposed by Xiang et al. with a precision and sensitivity of 99.91% and 99.77%, respectively over the MIT-DB [37]. Oh et al. tried to use UNET for "sample-wise classification" of ECG into R and Non-R peak labels. A major drawback of this proposed approach is that it has suffered too many false positives, caused by the misclassification of surrounding samples at the R peak, with a sensitivity level as low as 29.55% despite the fact that ECG quality is high [38].

All of the abovementioned algorithms were only tested on the high-quality clinical ECG records such as in the MIT-DB with a limited number of beats, i.e., the total number of beats of about 100K. A faster regional CNN was proposed by Yang et al. by turning a 1D ECG signal into a 2D image. The model was tested on the 24 hours wearable ECG recordings and showed promising results with 98.52% positive predictively and 98.76% sensitivity [39]. This was the first study that evaluated noisy and low-quality ECG records acquired by a wearable ECG sensor. Another major issue in using deep learning methods for R-peak detection is the limited number of ECG beats partitioned further for training, validation, and test sets. Besides, most of these methods test their robustness by artificially adding noise, i.e., baseline wanders and motion artifacts to ECG records in the MIT-DB. Such artificial artifact creation may not represent the actual variations and degradations that occur in Holter and mobile ECG devices. Thus, the performance of these algorithms would deteriorate in practice specifically on Holter devices where the signal is corrupted with a high level of noise and frequent artifacts and ECG baseline level varies drastically and abruptly. The challenges for accurate peak detection in Holter devices include powerline interference, baseline wandering, amplitude variability, multi-source PVC, noise, and other abnormal interference such as atrial fibrillation (AF) and electrode sliding interference. **Figure 1** shows typical ECG signals from the CPSC-DB where the benchmark Pan-Tompkins method yields many false-positives (FPs) and false-negatives (FNs). Therefore, there is a need for a robust and highly accurate R-peak detector for low-quality ECG signals.

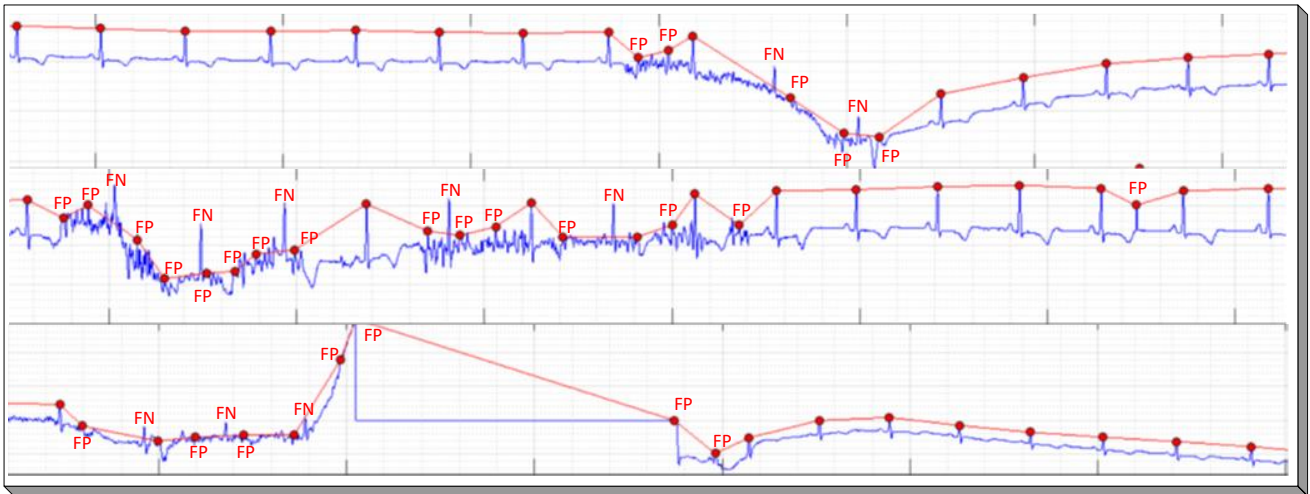

**Figure 1: Typical Holter ECG segments from the record of patient 6 in the CPSC dataset. Red circles represent the R-peaks detected by the Pan-Tompkins method where several false-positives (FPs) and false-negatives (FNs) are visible.**

To address the aforementioned limitations and drawbacks, in this study, we propose a novel and robust peak detection technique that can specifically be used for Holter devices or any other ECG acquisition system in non-clinical settings. For this purpose, first, we labeled the R-peak locations of the largest ECG Holter dataset with more than 1M beats. Then, we formulate R-peak detection as a 1D segmentation problem to enable the detection of the precise localization of R peaks with minimal post-processing. To accomplish this, the 1D Encoder-Decoder CNN model is constructed for the peak detection in a 20-second normalized ECG segment that outputs a reconstructed signal of equal length with a 1D segmentation map where a 5-sample pulse is centered at the location of each R-peak. It is also desirable to minimize (or suppress) false alarm rate especially when patients are in the intensive care unit (ICU) as higher false alarms decrease the quality of care by increasing patient delirium through noise pollution and slowing staff response times [19], [40], [41]. Thus, the proposed approach further improves the accuracy with a verification model that can approximately reduce 21% of the false alarms.

We evaluate the proposed approach on both benchmark datasets, MIT-DB and CPSC-DB, and thus, this becomes the first method that has ever been evaluated with more than 1M annotated beats (N, S, and V type beats). An extensive set of experimental results show that the proposed approach outperforms all *state-of-the-art* R-peak detection methods in CPSC-DB with a significant performance gap while it achieves a similar or better result in MIT-DB *without* explicit training in this dataset. The proposed method is further evaluated specifically for the R-peak detection of abnormal beats (e.g., S and V beats). This is indeed a challenging task compared to the detection of normal beats. In short, the novelty and significant contributions of the study are as follows:

- We propose a robust algorithm for the R-peak detection in low-quality Holter ECGs. Despite numerous classification methods in this domain, we approach this as a regression problem for utmost robustness and detection performance. From each ECG segment, a pulse train is produced from which the R-peaks can be detected through minimal post-processing.
- We proposed a novel verification model to reduce the number of false alarms in R peak detection with a significant margin.
- We demonstrated the generalization capacity of the proposed algorithm by evaluating it on the gold benchmark MIT dataset and achieved a *state-of-the-art* performance level without even training over this dataset.
- We provided ground truth peak locations for the largest Holter ECG dataset with more than 1M beats and this dataset will be released with R-peak annotations.
- Finally, this is the first study that has been quantitatively evaluated over the largest Holter ECG dataset for R peak detection. In particular, the proposed algorithm achieved enhanced sensitivity performance for detecting abnormal S and V beats specifically.

The rest of the paper is organized as follows: Section II outlines the ECG datasets used in this study. The detailed R-peak detection approach is presented in Section III. In Section IV, the performance and robustness of the proposed approach are evaluated over the two benchmark datasets using the standard performance metrics and the results are compared with the previous state-of-the-art works. Finally, Section V concludes the paper and suggests topics for future work.

## II. DATASETS

This section talks in detail about the two databases that are used for training and evaluation of the novel R-peak detection algorithm.

### A. China Physiological Signal Challenge-2020

The China Physiological Signal Challenge (2020) dataset (CPSC-DB) consists of 10 single-lead ECG recordings which are collected from arrhythmia patients, each of the recordings lasts for about 24 hours (shown in **Table I**) [42].

**Table I: Detailed information on the ECG data from CPSC-DB.**

| Patient | AF? | Length(h) | # Tot. Beats | # V Beats | # S Beats |
|---|---|---|---|---|---|
| 1 | No | 25.89 | 109731 | 0 | 24 |
| 2 | Yes | 22.83 | 108297 | 4554 | 0 |
| 3 | Yes | 24.70 | 138878 | 382 | 0 |
| 4 | No | 24.51 | 101734 | 19024 | 3466 |
| 5 | No | 23.57 | 94635 | 1 | 25 |
| 6 | No | 24.59 | 770806 | 0 | 6 |
| 7 | No | 23.11 | 96814 | 15150 | 3481 |
| 8 | Yes | 25.46 | 125495 | 2793 | 0 |
| 9 | No | 25.84 | 89854 | 2 | 1462 |
| 10 | No | 23.64 | 82851 | 169 | 9071 |

**Table I** also indicates if the patient has undergone atrial fibrillation (AF) or not. All ECG data were acquired by a unified wearable ECG device with a sampling frequency of 400 Hz and the total number of beats is 1,026,095. The recordings include irregular heart rhythms as well as SPB (S) and PVC (V) type beats. All recordings are provided in MATLAB format with corresponding S and V beats annotations. R-peak annotations for each ECG cycle were annotated by a team of biomedical researchers. To show the robustness of the R-peak detector against noise and other artifacts, CPSC-DB presents a real-world Holter dataset containing numerous ECG containments and artifacts.

*B. MIT-BIH arrhythmia dataset*

The second benchmark dataset used for evaluation and performance comparisons is the MIT-BIH arrhythmia dataset that consists of 48 two-lead ECGs from 47 subjects which are sampled at 360 Hz and each record covers 30 minutes [43]. This dataset has been widely used as a benchmark and it is the most popular clinical ECG database for the evaluation of peak detection algorithms. The signals in this dataset are relatively clean and of high-quality, collected in clinical settings. The modified lead II ECG was used in this study and the total number of beats is 109,475.

## III. METHODOLOGY

The proposed method is illustrated in **Figure 2**. The raw ECG segment of 20 seconds is fed into the proposed 1D CNN model. The model's output is further verified by another (verification) model that detects the false alarms. Unlike other approaches [35][36], all beats including arrhythmia beats were used for training and detection as the detection of arrhythmic beats is the most challenging problem.

*A. Problem formulation*

Encoder-Decoder (E-D) CNN models are popularly used in biomedical image segmentation applications. The model consists of a contracting as well as an expanding path. The contracting path consists of the repeated application of convolutions each followed by a Rectified Linear Unit (ReLU) and max pooling operation for downsampling. In contrast, the expanding path consists of the feature map upsampling, which is followed by a convolution ("up-convolution") after concatenation with the corresponding feature map from the contracting path, each followed by a ReLU.

We formulated peak detection as a 1D segmentation problem to segment R-peaks that involve signal to signal transformation and maps the input ECG to a pulse train where each pulse has a width of 5 samples and is centered at the R-peak location. The dataset consists of the $i^{th}$ input ECG signal, $x(i)$ linearly normalized between +1 and -1, and a 1D segmentation map, $y(i)$ that was obtained from R-peak locations. The proposed network is based on the implementation of the 1D convolutional encoder-decoder model. The encoder block takes a 20-second ECG segment as input, downsamples it after passing through convolutional kernels, and produces the compressed discriminative feature representation, $z(i)$. This compressed feature vector is then mapped reversely using the decoder block that upsamples the obtained features from the encoder to construct the output segmentation map $\hat{y}^{(i)}$ using Eq. (1).

$$z^{(i)} = E(x^{(i)}, \theta_1)$$
$$\hat{y}^{(i)} = D(z^{(i)}, \theta_2) \quad (1)$$

where E and D represent the encoder and decoder blocks. The weight vectors $\theta_1$ and $\theta_2$ in Eq. (1) are optimized by minimizing the binary cross-entropy (BCE) loss between the predicted and the actual R peak segmentation map.

$$Loss = \frac{1}{N}\sum_{i=1}^{N} y_i \times \log \hat{y} + (1 - y_i) \times \log(1 - y_i) \quad (2)$$

*B. Data Augmentation*

Data augmentation is an essential step when dealing with limited training examples, to teach the network the desired robustness and invariance. In the case of the CPSC dataset, arrhythmia beats are significantly scarcer than the normal beats. To make the S- (Supraventricular ectopic beat) and V- (Ventricular ectopic beat) beat detection more accurate and robust, we generate augmented arrhythmic beats from the 20 second ECG segments containing one or more arrhythmia beats by adding baseline wander and motion artifacts from the Noise Stress Test Database (NST-DB)[23]. The effect of augmentation on detector performance specifically for S- and V-beats has been discussed in detail in Section IV. Further implicit data augmentation is done by using Dropout layers at the end of the encoding layers.

*C. Network Architecture*

The architecture of the proposed model is adapted from the UNet model [44], which is commonly used in 2D medical image segmentation. 1D E-D CNN models were used both with and without skip connections. To show the effectiveness of skip connections that is the core idea in the UNet architecture, we also performed experiments without the skip connections. The proposed architecture of the model is illustrated in **Figure 2**. The proposed model is based on multiple hidden layers to learn high-level feature representations of ECG data.

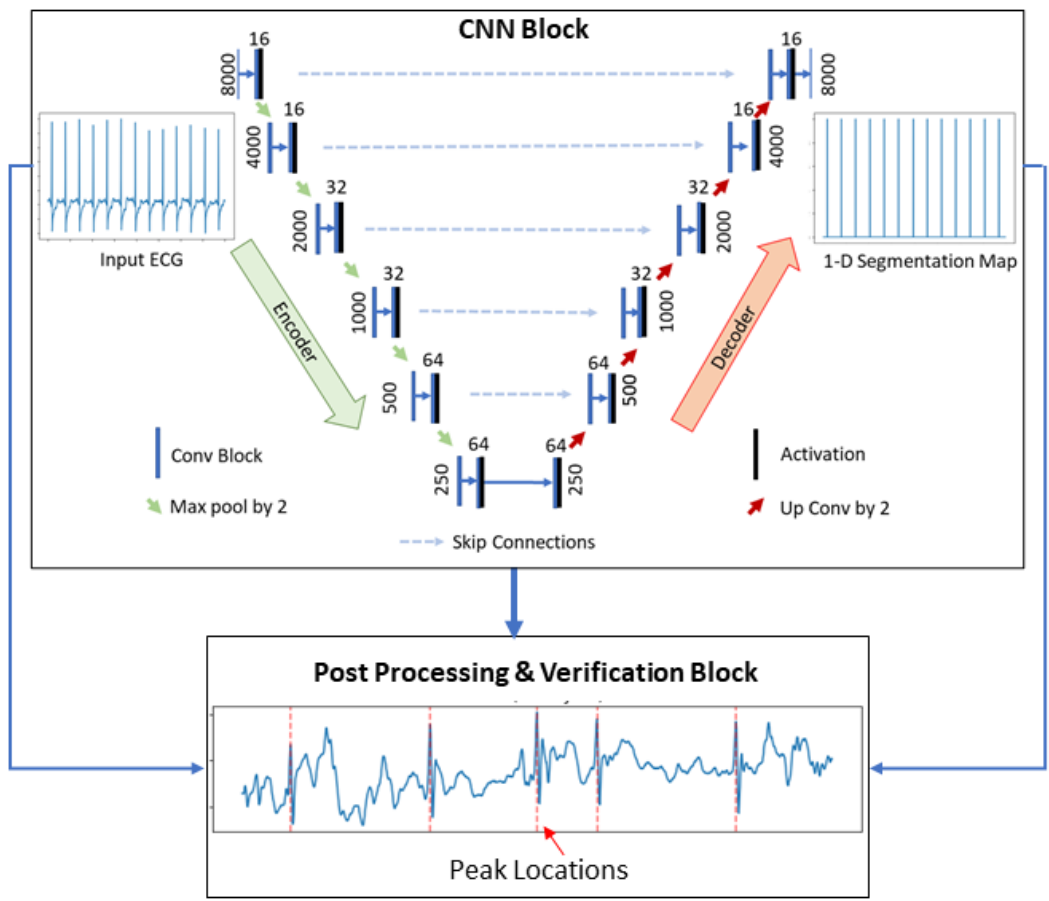

**Figure 2: The proposed systematic approach and network architecture.**

The input layer of both models accepts input with variable sizes. We chose 20 second ECG windows for the selected model architecture and parameters. However, the proposed method can be performed in any practical (e.g., 5s – 30s) window size. The model is designed for signal-to-signal mapping and it contains 6 layers in the encoder block as well as 6 layers in the decoder block with 1 output layer having a total of 38,209 trainable parameters. The input is downsampled by the encoder through 6 layers of convolutions with a downsampling factor of 2. The kernel size is set to be 9, 6, and 3 for every two consecutive layers. The number of filters starting from 16 is increased by a factor of two after every two consecutive layers. Each convolution is followed by batch normalization and Relu activation. In the decoder block, the compressed feature vector is upsampled by the same number of transpose convolutions with a reverse configuration of the encoder block.

The two main reasons to employ the batch normalization layer after the convolution layer are: improved generalization and speeding up the training process. To down-sample the 1D feature map while retaining important information, the max-pooling layer connected to the batch normalization layer is used. The features were downsampled by a factor of 2 in each encoding layer. Finally, the output 1-D segmentation mask is generated by using a filter size of 1 for convolution operation in the last layer. All convolutional layers in this architecture use the ReLU activation function, except for the final one. The last layer uses SoftMax activation to get a one-dimensional segmentation map for R peaks.

### D. Training

The proposed CNN model for each fold is implemented in Python by using Keras which is the high-level API of TensorFlow, a highly productive platform for solving machine problems. To train both variants (UNET and E-D CNN), Adam optimizer is used with a learning rate of $10^{-3}$, and the network parameters of the model were initialized randomly in the range of [-0.1, 0.1].

The models are trained for 50 epochs with a batch size of 64. 10-folds cross-validation is used, i.e., for each fold ECG records of the 9 patients were used to train the model, and the remaining one was used for evaluation. Data augmentation techniques over the abnormal beats such as adding Gaussian noise with random variance, combining a sinusoidal signal with random initial phase and amplitude were used. Finally, as the objective function for training the network, we used the cross-entropy loss [2] and the loss is summed up over all the available samples in a mini-batch.

### E. Verification Model

Once R-peak locations with their occurrence probabilities have been obtained from the 1D segmentation map, they are then passed through a verification model to remove any unlikely R-peak based on a timing criterion, i.e., if the predicted locations of the two beats fall within a window of 300 milliseconds, one of the beats is identified as a false alarm. The window length of 300 milliseconds is chosen to avoid the removal of actual beats as FPs in AF patients. In AF, heart rate can go up to 125-150 beats per minute due to the rapid rate of fibrillatory impulses [45]. Firstly, to capture the morphology of the whole QRS complex, 60 samples window is extracted from both beats by taking the predicted R-peak location as the midpoint of the beat. Then, the two reference beats of the same sample size are extracted. The criteria for choosing these reference beats is that the probability of the beats must be greater than 0.5. The first beat that satisfies this criterion before and after two closely located beats are selected. The beat similarity score ($S_{B_i}$) for the consecutive beats, $i = 1,2$ is expressed by the formula as follows:

$$S_{B_i} = \rho(B_i, B_P) \times \rho(B_i, B_N) \quad (4)$$

where $B_P$ and $B_N$ are previous and next reference beats, respectively. The Pearson product-moment correlation coefficient ($\rho$) can be expressed as,

$$\rho_{X,Y} = \frac{cov(X,Y)}{\sigma_X \times \sigma_Y} \quad (5)$$

where $cov$ is covariance and $\sigma$ is the standard deviation.

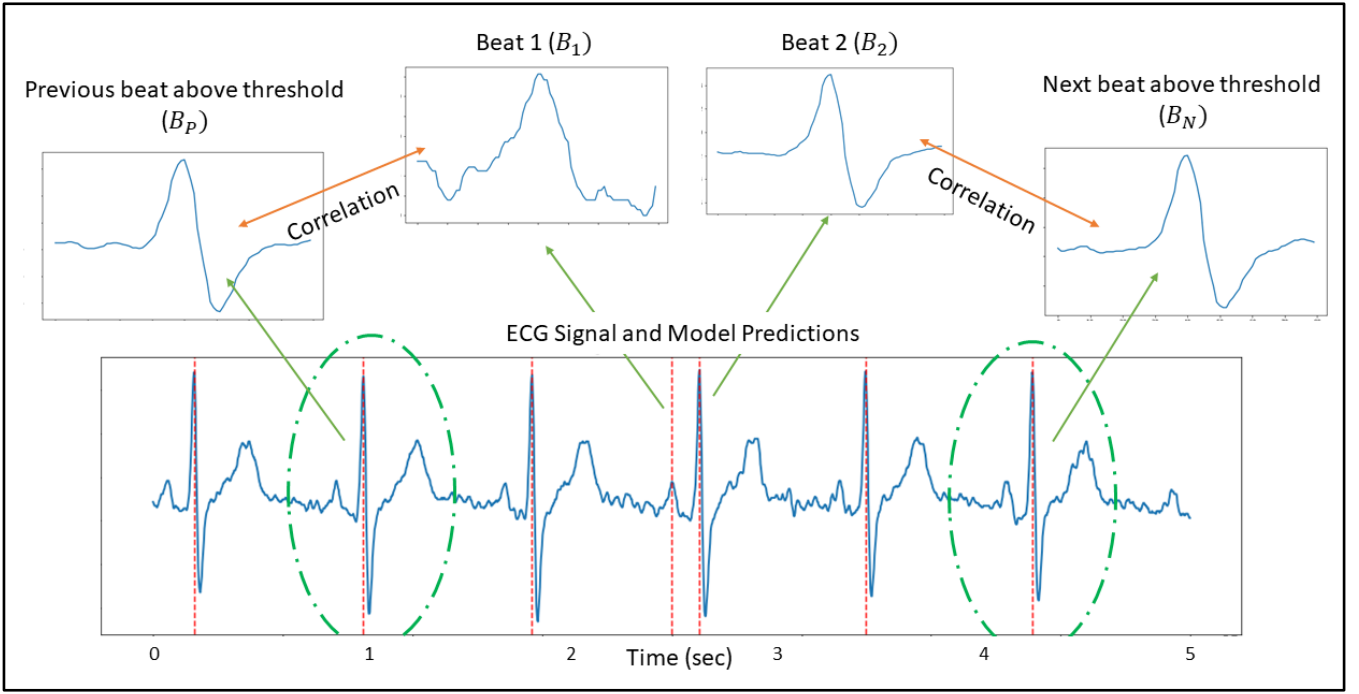

**Figure 3: Illustration of the verification model. Red dotted lines represent the R peak predictions of the proposed model. Two closely predicted beats ($B_1 and B_2$) are visible. $B_1$ will be detected as false peak by the verification model due to a low beat score and it will be removed automatically.**

The beat with a minimum similarity score ($S_{B_i}$) is assumed to be a false beat and hence, removed from the R-peak locations. The verification model is illustrated in **Figure 3**.

## IV. EXPERIMENTAL RESULTS

In this section, we will first present the experimental setup used for testing and evaluation of the proposed R-peak detector on 1D E-D CNN. An extensive set of peak detection experiments and comparative evaluations against the five *state-of-the-art* classifiers from the literature over both datasets will be presented next. For CPSC-DB we performed 10-fold cross-validation. For MIT-DB, the proposed model was trained only on the CPSC-DB data and tested on MIT-DB. For the proposed method, ECG records from MIT-DB were resampled at 400 Hz to match the sampling frequency of the training data (CPSC-DB).

### A. Evaluation metrics

Quantitative performance of the different models and approaches are compared using three performance measures that are commonly used: Precision, Recall, and F1-score. Note that, the measurement of True Positives (TP), False Negatives (FN), and False Positives (FP) were taken within a tolerance of ±75 ms [2] of the truth peak location.

$$Recall\ (\%) = \frac{TP}{TP + FN} \times 100 \quad (4)$$

$$Precision\ (\%) = \frac{TP}{TP + FP} \times 100 \quad (5)$$

$$F1\ (\%) = \frac{2\ \times Precision \times Recall}{Precision + Recall} \times 100 \qquad (6)$$

Since this is an R-peak detection operation, True Negatives (TN) do not exist as a performance measure.

### B. Results on CPSC-DB

**Table II** exhibits the R-peak detection performance of CPSC-DB. From the results, it is evident that the proposed model achieved the top performance on CPSC-DB with a top F1 score of 99.30%. Moreover, the proposed approach is very effective in detecting S and V type beats as it only misses less than 1% of total arrhythmia beats while most of the competing methods do not even consider the arrhythmia beats in their experiments. Compared to all competing methods, the proposed approach reduced the FPs and FNs on R-peak detection by more than 54% and 82%, respectively. Such a substantial reduction especially on FNs over all competing methods shows that the proposed approach with 1D CNNs can indeed accomplish a superior learning capability and robustness to detect the actual peaks on such low-quality ECG signal.

The peak location of S and V beats is crucial because peak detection is the prior operation to an automated ECG beat classification and arrhythmia detection. This aim cannot be fulfilled if the peak detector fails on an abnormal beat. As shown in **Table II**, once again the proposed approach has achieved a significant performance gap on detecting both S and V beats over the competing methods, i.e., the proposed approach reduced the missed S and V beats (FNs) by more than 80% and 95%, respectively.

On the other hand, the poor performance of the competing models is because most of them are designed for high-quality clinical ECGs, whereas several artifacts and other variations present in the Holter ECG database significantly deteriorated their performances. There are also other morphological variations in Holter ECGs that are not present in ECGs acquired in clinical settings.

### C. Results on MIT-DB

The proposed 1D CNN model with skip connections (UNET) was trained over the 10 records of CPSC-DB and was solely evaluated on MIT-DB. The results presented in **Table III** show that the proposed model achieves better detection performance than the state-of-the-art methods except one where the performance gap is insignificant. Such a "similar or better" performance is achieved even though the proposed model is trained only on CPSC-DB data while all the MIT-DB ECG records were unseen and used solely for testing.

The contribution of data augmentation and verification on the detection performance can be seen from the results presented in Table IV. By data augmentation, we were able to reduce 40% of missed detections (FNs). The tradeoff for this reduction is the 9% increase in the number of false alarms (FPs). As a solution, by applying the verification model, 21% of the false alarms were reduced and the state-of-the-art performance level is achieved.

**Table II: Peak detection performances of the proposed approaches and competing algorithms on CPSC-DB. The best results are highlighted in bold.**

| Methods | TP | FN | FP | Recall | Precision | F1 | S Missed | V missed |
|---|---|---|---|---|---|---|---|---|
| **Proposed (UNET)** | **1,022,845** | **3,250** | **10,916** | **99.69** | **98.91** | **99.30** | **40** | **552** |
| **Proposed (E-D CNN)** | 1,021,868 | 4,227 | 12,225 | 99.61 | 98.75 | 99.17 | 204 | 1,057 |
| LSTM [35] | 1,007,823 | 18,272 | 23,835 | 98.20 | 97.76 | 97.88 | 205 | 12,828 |
| P and T [46] | 998,413 | 27,682 | 28,940 | 97.31 | 97.18 | 97.23 | 305 | 10,706 |
| Hamilton [47] | 993,920 | 32,175 | 62,733 | 96.82 | 93.73 | 95.14 | 372 | 10,169 |
| Two Moving Avg. [48] | 992,305 | 33,790 | 46,349 | 96.66 | 95.34 | 95.97 | 304 | 13,103 |
| SWT [49] | 975,222 | 50,873 | 27,936 | 95.06 | 97.20 | 96.09 | 517 | 14,792 |

**Table III: Peak detection performances of the proposed approaches and competing algorithms on MIT-DB. The best results are highlighted in bold.**

| Methods | Beats | TP | FN | FP | Recall | Precision | F1 |
|---|---|---|---|---|---|---|---|
| **Proposed (UNET)** | 109,475 | 109,304 | 171 | **182** | 99.85 | 99.82 | 99.83 |
| P and T [46] | 109,809 | 109,208 | 507 | 601 | 99.45 | 99.54 | 99.50 |
| Hamilton [47] | 109,267 | 108,927 | 248 | 340 | 99.69 | 99.77 | 99.73 |
| Two Moving Avg. [48] | 109,493 | 109,397 | **97** | 1,715 | 98.31 | **99.92** | 99.11 |
| SWT [49] | | | | | **99.88** | 99.84 | **99.86** |

Table IV: Effect of augmentation and verification on the proposed approach. The best results are highlighted in bold.

| Model | TP | FN | FP | Recall | Precision | F1 | S missed | V missed |
|---|---|---|---|---|---|---|---|---|
| No Aug | 1,020,679 | 5,416 | 12,764 | 99.5 | 98.77 | 99.13 | 87 | 1,760 |
| Aug | **1,022,874** | **3,221** | 13,931 | 99.7 | 98.63 | 99.16 | **40** | **549** |
| Aug + Ver | 1,022,845 | 3,250 | **10,916** | 99.69 | **98.91** | **99.3** | **40** | 552 |

Some typical ECG records from different patients and their corresponding R-peaks detected by the proposed approach are shown in **Figure 4**. As in the cases shown in **Figure 1**, it is hard for an untrained human eye to detect some of the R-peaks accurately. Even though extreme baseline variations, structural artifacts, and noise are present, the proposed approach can detect almost all R-peaks accurately. Finally, it is worth mentioning that the R-peak locations perfectly align with the ground truth positions.

*D. Computation Time*

We implemented the proposed peak detection algorithm using Python and Keras machine learning libraries. All the experiments reported in this paper were performed on a 2.2GHz Intel Core i7-8750H with 16 GB of RAM and an NVIDIA GeForce GTX 1060 graphic card. The training phase of the model was processed by CUDA kernels whereas the testing phase is implemented without a parallelized (CUDA) version with a single CPU. It takes about 191 msec to compute the segmentation mask for a 20s ECG segment. P&T method takes around 73 msec to process an ECG segment of the same length. The verification model only takes 11 msec to validate the detected peaks for false detections. The most important advantage of the proposed system is its fast processing for beat detection. Specifically, for the single-CPU implementation, the total time for a 20s ECG segment of a signal to obtain the peak locations is about 202 msec and this indicates around 100-times the real-time speed. This might be slower when using in Holter devices with a low-configuration processor, but it will still be several times faster than real-time processing speed.

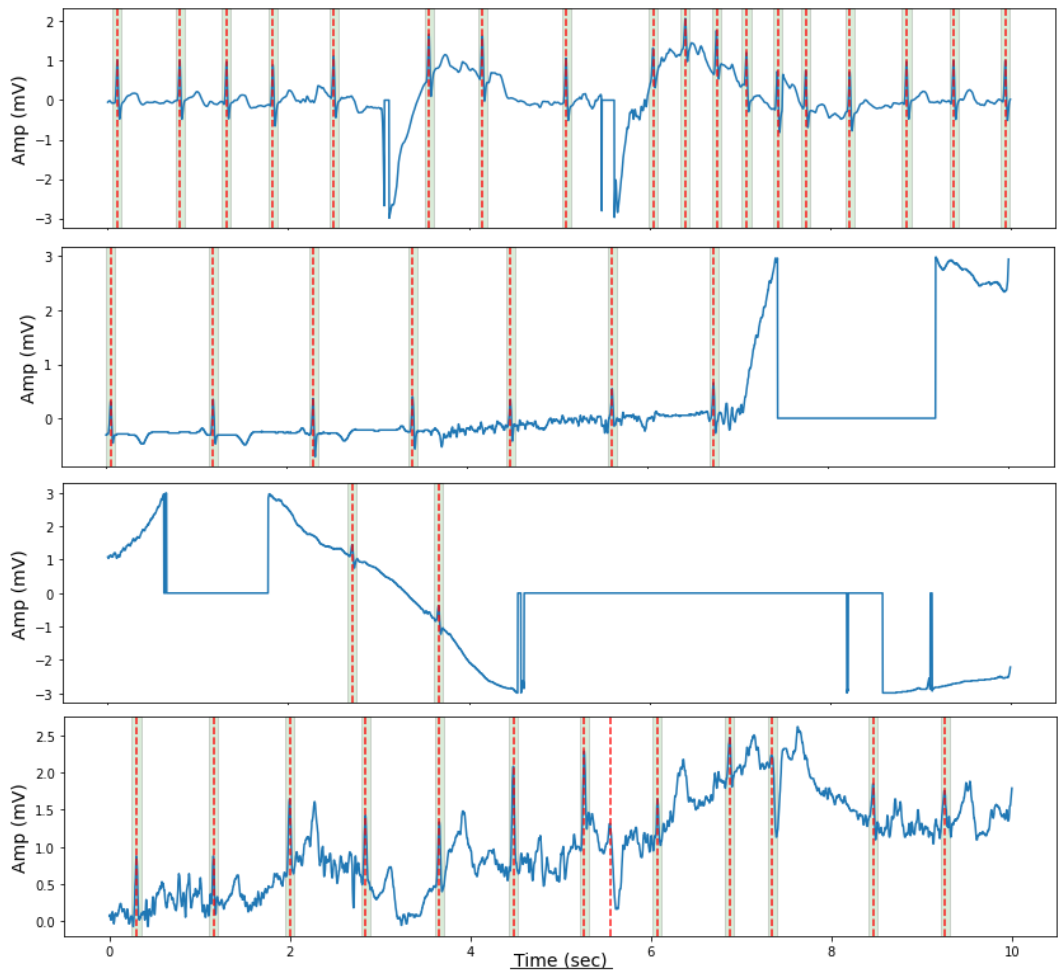

**Figure 4:** Some typical R-peak detection results of the proposed algorithm. Green lines represent the ground truth location of R-peaks and red dotted lines represent the peaks detected by the model.

## V. Conclusion

This study describes a novel approach for R-peak detection especially in low-quality Holter ECGs using 1-D CNN along with a verification model. The R-peak detection is approached as a 1D segmentation problem; hence the R-peaks are detected with minimal post-processing. We performed 10-fold comparative evaluations over the benchmark CPSC-DB with more than 1M beats. Against the competing *state-of-the-art* methods, the proposed approach not only achieved the highest detection performances, but it further reduced both FPs and FNs significantly. The most crucial advantage over the competing methods is that the proposed approach can reduce the number of *missed* S and V beats (FNs) by more than 80% and 95%, respectively. Over the MIT-DB dataset with high-quality ECG records, it also achieves similar or better performance than the competitors. Finally, since the proposed 1D CNN model performs only 1D convolutions, it achieved 100-times of the real-time speed in a standard computer, and thus, especially for low-power, mobile devices such as Holter monitors, the proposed approach can conveniently be used as an R-peak detector in real-time. In future work, we aim to explore robust quantization and model compression techniques to further reduce the model parameters and accelerate the inference process. Using a new-generation heterogeneous network model such as 1D Operational Neural Networks [50]–[54] instead of 1D CNNs is also planned to further improve the performance and to reduce the complexity.

## Acknowledgment

This work was supported by the Qatar National Research Fund (QNRF) through the ongoing project under Grant NPRP11S-0108-180228 and the Academy of Finland, project AWcHA.

## References

[1] J. Behar, J. Oster, Q. Li, and G. D. Clifford, "ECG signal quality during arrhythmia and its application to false alarm reduction," *IEEE Trans. Biomed. Eng.*, vol. 60, no. 6, pp. 1660–1666, 2013, doi: 10.1109/TBME.2013.2240452.

[2] F. Liu *et al.*, "Performance Analysis of Ten Common QRS Detectors on Different ECG Application Cases," *J. Healthc. Eng.*, vol. 2018, 2018, doi: 10.1155/2018/9050812.

[3] J. Behar, J. Oster, Q. Li, and G. D. Clifford, "ECG signal quality during arrhythmia and its application to false alarm reduction," *IEEE Trans. Biomed. Eng.*, vol. 60, no. 6, pp. 1660–1666, 2013, doi: 10.1109/TBME.2013.2240452.

[4] Q. Li, R. G. Mark, and G. D. Clifford, "Robust heart rate estimation from multiple asynchronous noisy sources using signal quality indices and a Kalman filter," *Physiol. Meas.*, vol. 29, no. 1, pp. 15–32, Jan. 2008, doi: 10.1088/0967-3334/29/1/002.

[5] W. J. Tompkins, "A Real-Time QRS Detection Algorithm," 1985.

[6] P. S. Hamilton and W. J. Tompkins, "Quantitative Investigation of QRS Detection Rules Using the MIT/BIH Arrhythmia Database," *IEEE Trans. Biomed. Eng.*, vol. BME-33, no. 12, pp. 1157–1165, 1986, doi: 10.1109/TBME.1986.325695.

[7] M. Jia, F. Li, J. Wu, Z. Chen, and Y. Pu, "Robust QRS Detection Using High-Resolution Wavelet Packet Decomposition and Time-Attention Convolutional Neural Network," *IEEE Access*, vol. 8, pp. 16979–16988, 2020, doi: 10.1109/ACCESS.2020.2967775.

[8] D. S. Benitez, P. A. Gaydecki, A. Zaidi, and A. P. Fitzpatrick, "New QRS detection algorithm based on the Hilbert transform," in *Computers in Cardiology*, 2000, pp. 379–382, doi: 10.1109/cic.2000.898536.

[9] N. M. Arzeno, Z. De Deng, and C. S. Poon, "Analysis of first-derivative based QRS detection algorithms," *IEEE Trans. Biomed. Eng.*, vol. 55, no. 2, pp. 478–484, Feb. 2008, doi: 10.1109/TBME.2007.912658.

[10] A. Martínez, R. Alcaraz, and J. J. Rieta, "Application of the phasor transform for automatic delineation of single-lead ECG fiducial points," *Physiol. Meas.*, vol. 31, no. 11, pp. 1467–1485, Nov. 2010, doi: 10.1088/0967-3334/31/11/005.

[11] B. Hossain, S. Khairul Bashar, A. J. Walkey, D. D. Mcmanus, and K. H. Chon, "An Accurate QRS Complex and P Wave Detection in ECG Signals Using Complete Ensemble Empirical Mode Decomposition with Adaptive Noise Approach," doi: 10.1109/ACCESS.2019.2939943.

[12] Z. E. Hadj Slimane and A. Naït-Ali, "QRS complex detection using Empirical Mode Decomposition," *Digit. Signal Process. A Rev. J.*, vol. 20, no. 4, pp. 1221–1228, Jul. 2010, doi: 10.1016/j.dsp.2009.10.017.

[13] V. Kalidas and L. Tamil, "Real-time QRS detector using stationary wavelet transform for automated ECG analysis," 2017, doi: 10.1109/BIBE.2017.00-12.

[14] S. Sahoo, B. Kanungo, S. Behera, and S. Sabut, "Multiresolution wavelet transform based feature extraction and ECG classification to detect cardiac abnormalities," *Meas. J. Int. Meas. Confed.*, vol. 108, pp. 55–66, Oct. 2017, doi: 10.1016/j.measurement.2017.05.022.

[15] S. Pal and M. Mitra, "Empirical mode decomposition based ECG enhancement and QRS detection," *Comput. Biol. Med.*, vol. 42, no. 1, pp. 83–92, Jan. 2012, doi: 10.1016/j.compbiomed.2011.10.012.

[16] M. A. Kabir and C. Shahnaz, "Denoising of ECG signals based on noise reduction algorithms in EMD and wavelet domains," *Biomed. Signal Process. Control*, vol. 7, no. 5, pp. 481–489, Sep. 2012, doi: 10.1016/j.bspc.2011.11.003.

[17] A. J. Nimunkar and W. J. Tompkins, "R-peak detection and signal averaging for simulated stress ECG using EMD," in *Annual International Conference of the IEEE Engineering in Medicine and Biology - Proceedings*, 2007, pp. 1261–1264, doi: 10.1109/IEMBS.2007.4352526.

[18] H. Rabbani, M. Parsa Mahjoob, E. Farahabadi, and A. Farahabadi, "R peak detection in electrocardiogram signal based on an optimal combination of wavelet

transform, Hilbert transform, and adaptive thresholding," *J. Med. Signals Sens.*, vol. 1, no. 2, pp. 91–98, May 2011, doi: 10.4103/2228-7477.95292.

[19] G. D. Clifford *et al.*, "False alarm reduction in critical care," *Physiological Measurement*, vol. 37, no. 8. Institute of Physics Publishing, pp. E5–E23, 2016, doi: 10.1088/0967-3334/37/8/E5.

[20] S. Parsons and J. Huizinga, "Robust and fast heart rate variability analysis of long and noisy electrocardiograms using neural networks and images," *arXiv*, Feb. 2019, Accessed: Dec. 10, 2020. [Online]. Available: http://arxiv.org/abs/1902.06151.

[21] M. Elgendi, "Fast QRS Detection with an Optimized Knowledge-Based Method: Evaluation on 11 Standard ECG Databases," *PLoS One*, vol. 8, no. 9, p. 73557, Sep. 2013, doi: 10.1371/journal.pone.0073557.

[22] M. Elgendi, B. Eskofier, S. Dokos, and D. Abbott, "Revisiting QRS detection methodologies for portable, wearable, battery-operated, and wireless ECG systems," *PLoS One*, vol. 9, no. 1, p. e84018, Jan. 2014, doi: 10.1371/journal.pone.0084018.

[23] Moody GB, Muldrow WE, and Mark RG, "The MIT-BIH Noise Stress Test Database," in *Computers in Cardiology*, 1984, pp. 381–384, doi: 10.13026/C2HS3T.

[24] B. Porr and L. Howell, "R-peak detector stress test with a new noisy ECG database reveals significant performance differences amongst popular detectors," *bioRxiv*, p. 722397, Aug. 2019, doi: 10.1101/722397.

[25] K. Arbateni and A. Bennia, "Sigmoidal radial basis function ANN for QRS complex detection," *Neurocomputing*, vol. 145, pp. 438–450, Dec. 2014, doi: 10.1016/j.neucom.2014.05.009.

[26] A. S. A. Huque, K. I. Ahmed, M. A. Mukit, and R. Mostafa, "HMM-based Supervised Machine Learning Framework for the Detection of fECG R-R Peak Locations," *IRBM*, vol. 40, no. 3, pp. 157–166, Jun. 2019, doi: 10.1016/j.irbm.2019.04.004.

[27] G. Vijaya, V. Kumar, H. K. Verma, G. Viiayat, V. Kumarf, and H. K. Verma$, "ANN-based QRS-complex analysis of ECG A"-based QRS-complex analysis of ECG*," *Technol.*, vol. 22, no. 4, pp. 160–167, 1998, doi: 10.3109/03091909809032534.

[28] R. Rodríguez, A. Mexicano, J. Bila, S. Cervantes, and R. Ponce, "Feature extraction of electrocardiogram signals by applying adaptive threshold and principal component analysis," *J. Appl. Res. Technol.*, vol. 13, no. 2, pp. 261–269, Apr. 2015, doi: 10.1016/j.jart.2015.06.008.

[29] S. Kiranyaz, T. Ince, O. Abdeljaber, O. Avci, and M. Gabbouj, "1-D Convolutional Neural Networks for Signal Processing Applications," in *ICASSP, IEEE International Conference on Acoustics, Speech and Signal Processing - Proceedings*, May 2019, vol. 2019-May, pp. 8360–8364, doi: 10.1109/ICASSP.2019.8682194.

[30] S. Kiranyaz, O. Avci, O. Abdeljaber, T. Ince, M. Gabbouj, and D. J. Inman, "1D convolutional neural networks and applications: A survey," *Mech. Syst. Signal Process.*, vol. 151, Apr. 2021, doi: 10.1016/j.ymssp.2020.107398.

[31] T. Ince, S. Kiranyaz, L. Eren, M. Askar, and M. Gabbouj, "Real-Time Motor Fault Detection by 1-D Convolutional Neural Networks," *IEEE Trans. Ind. Electron.*, vol. 63, no. 11, pp. 7067–7075, Nov. 2016, doi: 10.1109/TIE.2016.2582729.

[32] S. Kiranyaz, T. Ince, and M. Gabbouj, "Real-Time Patient-Specific ECG Classification by 1-D Convolutional Neural Networks," *IEEE Trans. Biomed. Eng.*, vol. 63, no. 3, pp. 664–675, Mar. 2016, doi: 10.1109/TBME.2015.2468589.

[33] S. Kiranyaz, T. Ince, and M. Gabbouj, "Personalized Monitoring and Advance Warning System for Cardiac Arrhythmias," *Sci. Rep.*, vol. 7, no. 1, Dec. 2017, doi: 10.1038/s41598-017-09544-z.

[34] X. Wang and Q. Zou, "QRS detection in ECG signal based on residual network," in *2019 IEEE 11th International Conference on Communication Software and Networks, ICCSN 2019*, Jun. 2019, pp. 73–77, doi: 10.1109/ICCSN.2019.8905308.

[35] J. Laitala *et al.*, "Robust ECG R-peak detection using LSTM," *Proc. ACM Symp. Appl. Comput.*, no. March, pp. 1104–1111, 2020, doi: 10.1145/3341105.3373945.

[36] S. Vijayarangan, R. Vignesh, B. Murugesan, P. Sp, J. Joseph, and M. Sivaprakasam, "RPnet: A Deep Learning approach for robust R Peak detection in noisy ECG," *Proc. Annu. Int. Conf. IEEE Eng. Med. Biol. Soc. EMBS*, vol. 2020-July, no. 1, pp. 345–348, 2020, doi: 10.1109/EMBC44109.2020.9176084.

[37] Y. Xiang, Z. Lin, and J. Meng, "Automatic QRS complex detection using two-level convolutional neural network," *Biomed. Eng. Online*, vol. 17, no. 1, pp. 1–17, 2018, doi: 10.1186/s12938-018-0441-4.

[38] S. L. Oh, E. Y. K. Ng, R. S. Tan, and U. R. Acharya, "Automated beat-wise arrhythmia diagnosis using modified U-net on extended electrocardiographic recordings with heterogeneous arrhythmia types," *Comput. Biol. Med.*, vol. 105, pp. 92–101, Feb. 2019, doi: 10.1016/j.compbiomed.2018.12.012.

[39] H. Yang, M. Huang, Z. Cai, Y. Yao, and C. Liu, "A Faster R CNN-based Real-time QRS Detector," in *2019 Computing in Cardiology Conference (CinC)*, Dec. 2019, vol. 45, doi: 10.22489/cinc.2019.053.

[40] N. Sadr, J. Huvanandana, D. T. Nguyen, C. Kalra, A. McEwan, and P. De Chazal, "Reducing false arrhythmia alarms in the ICU using multimodal signals and robust QRS detection," *Physiol. Meas.*, vol. 37, no. 8, pp. 1340–1354, Jul. 2016, doi: 10.1088/0967-3334/37/8/1340.

[41] L. M. Eerikäinen, J. Vanschoren, M. J. Rooijakkers, R. Vullings, and R. M. Aarts, "Decreasing the false alarm rate of arrhythmias in intensive care using a machine learning approach," in *Computing in Cardiology*, Feb. 2015, vol. 42, pp. 293–296, doi: 10.1109/CIC.2015.7408644.

[42] "Challenge Data - CPSC2020." http://2020.cpscsub.com/Data (accessed Dec. 29, 2020).

[43] "MIT-BIH Arrhythmia Database v1.0.0." https://physionet.org/content/mitdb/1.0.0/ (accessed

Dec. 29, 2020).
[44] O. Ronneberger, P. Fischer, and T. Brox, "U-net: Convolutional networks for biomedical image segmentation," in *Lecture Notes in Computer Science (including subseries Lecture Notes in Artificial Intelligence and Lecture Notes in Bioinformatics)*, 2015, vol. 9351, pp. 234–241, doi: 10.1007/978-3-319-24574-4_28.
[45] "Guyton A.C, Hall J.E. Textbook of Medical Physiology. 11th ed. Vol. 11. Philadelphia, PA, Pennsylvania: Elsevier, Saunders; 2006. - Google Search." .
[46] J. Pan and W. J. Tompkins, "A Real-Time QRS Detection Algorithm," *IEEE Trans. Biomed. Eng.*, vol. BME-32, no. 3, pp. 230–236, 1985, doi: 10.1109/TBME.1985.325532.
[47] P. S. Hamilton and E. P. Limited, "Open Source ECG Analysis Software Documentation," 2002. Accessed: Dec. 10, 2020. [Online]. Available: http://www.eplimited.com/.
[48] M. Elgendi, M. Jonkman, and F. De Boer, "Frequency bands effects on QRS detection." Institute for Systems and Technologies of Information, Control and Communication (INSTICC), pp. 428–431, 2010, Accessed: Dec. 10, 2020. [Online]. Available: https://researchers.cdu.edu.au/en/publications/frequency-bands-effects-on-qrs-detection.
[49] V. Kalidas, L. T.-2017 I. 17th I. Conference, and undefined 2017, "Real-time QRS detector using stationary wavelet transform for automated ECG analysis," *ieeexplore.ieee.org*, Accessed: Dec. 10, 2020. [Online]. Available: https://ieeexplore.ieee.org/abstract/document/8251332/?casa_token=t1JYffAAqCMAAAAA:mJABGVQwxiAJJUd7eDUbwyKgaj5BCS2LADWS6OLHrW01Y7VPzTPe1xBgJNk-bjelV5cmZkHbFQ.
[50] S. Kiranyaz, T. Ince, A. Iosifidis, and M. Gabbouj, "Operational neural networks," *Neural Comput. Appl.*, vol. 32, no. 11, pp. 6645–6668, Jun. 2020, doi: 10.1007/s00521-020-04780-3.
[51] S. Kiranyaz, J. Malik, H. Ben Abdallah, T. Ince, A. Iosifidis, and M. Gabbouj, "Exploiting Heterogeneity in Operational Neural Networks by Synaptic Plasticity," Aug. 2020, Accessed: Dec. 27, 2020. [Online]. Available: http://arxiv.org/abs/2009.08934.
[52] J. Malik, S. Kiranyaz, and M. Gabbouj, "FastONN -- Python based open-source GPU implementation for Operational Neural Networks," Jun. 2020, Accessed: Dec. 27, 2020. [Online]. Available: http://arxiv.org/abs/2006.02267.
[53] S. Kiranyaz, J. Malik, H. Ben Abdallah, T. Ince, A. Iosifidis, and M. Gabbouj, "Self-Organized Operational Neural Networks with Generative Neurons," *arXiv*, Apr. 2020, Accessed: Dec. 27, 2020. [Online]. Available: http://arxiv.org/abs/2004.11778.
[54] J. Malik, S. Kiranyaz, and M. Gabbouj, "Self-Organized Operational Neural Networks for Severe Image Restoration Problems," Aug. 2020, Accessed: Dec. 27, 2020. [Online]. Available: http://arxiv.org/abs/2008.12894.